\documentclass[sn-vancouver,icol]{sn-jnl}

\usepackage{lineno}

\jyear{2021}%

\theoremstyle{thmstyleone}%
\theoremstyle{thmstyletwo}%

\theoremstyle{thmstylethree}%

\begin{document}

\title[ ]{Differences between the true reproduction number and the apparent reproduction number of an epidemic time series}


\author*[1,2,3]{\fnm{Oliver} \sur{Eales}}\email{oliver.eales@unimelb.edu.au}
\author[2,3]{\fnm{Steven} \sur{Riley}}\email{s.riley@imperial.ac.uk}

\affil[1]{\orgdiv{Infectious Disease Dynamics Unit, Centre for Epidemiology and Biostatistics, Melbourne School of Population and Global Health}, \orgname{The University of Melbourne}, \orgaddress{ \city{Melbourne}, \country{Australia}}}

\affil[2]{\orgdiv{School of Public Health}, \orgname{Imperial College London}, \orgaddress{ \city{London}, \country{United Kingdom}}}

\affil[3]{\orgdiv{MRC Centre for Global infectious Disease Analysis and Abdul Latif Jameel Institute for Disease and Emergency Analytics}, \orgname{Imperial College London}, \orgaddress{ \city{London}, \country{United Kingdom}}}

\abstract{
The time-varying reproduction number $R(t)$ measures the number of new infections per infectious individual and is closely correlated with the time series of infection incidence by definition. The timings of actual infections are rarely known, and analysis of epidemics usually relies on time series data for other outcomes such as symptom onset. A common implicit assumption, when estimating $R(t)$ from an epidemic time series, is that $R(t)$ has the same relationship with these downstream outcomes as it does with the time series of incidence. However, this assumption is unlikely to be valid given that most epidemic time series are not perfect proxies of incidence. Rather they represent convolutions of incidence with uncertain delay distributions. Here we define the apparent time-varying reproduction number, $R_A(t)$, the reproduction number calculated from a downstream epidemic time series and demonstrate how differences between $R_A(t)$ and $R(t)$ depend on the convolution function. The mean of the convolution function sets a time offset between the two signals, whilst the variance of the convolution function introduces a relative distortion between them. We present the convolution functions of epidemic time series that were available during the SARS-CoV-2 pandemic. Infection prevalence, measured by random sampling studies, presents fewer biases than other epidemic time series. Here we show that additionally the mean and variance of its convolution function were similar to that obtained from traditional surveillance based on mass-testing and could be reduced using more frequent testing, or by using stricter thresholds for positivity. Infection prevalence studies continue to be a versatile tool for tracking the temporal trends of $R(t)$, and with additional refinements to their study protocol, will be of even greater utility during any future epidemics or pandemics.
}

\keywords{Reproduction number, Epidemics, SARS-CoV-2, Pandemics, COVID-19, Infection prevalence, Disease surveillance, Testing}

\maketitle

\section{Introduction}
Infectious disease epidemics are a major threat to public health. Over the past two decades there has been major outbreaks of SARS \cite{Anderson2004-dx}, MERS \cite{Sharif-Yakan2014-dg}, influenza \cite{Gog2014-fy}, Ebola \cite{Gire2014-de}, dengue \cite{MacCormack-Gelles2018-wk} and most recently the emergence of SARS-CoV-2 \cite{Li2021-wx}, which resulted in the COVID-19 pandemic. Accurately quantifying the transmission dynamics of infection is crucial for informing decisions about changes to policy on public health interventions.\\
\indent
The reproduction number, the expected number of secondary infections resulting from a typical primary infection, is a key quantity describing a pathogen’s transmission dynamics. If public health interventions are to be implemented, $R(t)$ can inform on the magnitude of the interventions required to bring the epidemic under control \cite{Anderson1992-xa}. Analysis of how $R(t)$ changes over time is also important for assessing the impact of interventions including vaccination and social distancing. \cite{Li2021-uz, Eales2022-ub}.\\
\indent  
Mathematically, $R(t)$ is linked to the time series of infection incidence (the rate of new infections) by a renewal process. If the infection incidence and the generation time (the time between a primary and secondary infection) distribution are known, then $R(t)$ can simply be calculated \cite{Wallinga2007-qv}. However, the timings of infections are rarely known, and so accurate time series of infection incidence rarely exist. Estimates of $R(t)$ must instead rely on other epidemic time series that are often treated as proxies for the infection incidence \cite{Nash2022-ob}.\\
\indent
During an epidemic there are often many sources of time series data collected. These can include measurements of the frequency of downstream points in the natural history of infection such as onset of symptoms, hospitalisation, and death \cite{noauthor_undated-pb}. During the SARS-CoV-2 pandemic, time series data for infection prevalence (the proportion of people testing positive for the virus) was also available in England \cite{Elliott2022-xj, Chadeau-Hyam2022-kc}. Infection prevalence time series are often less biased than other epidemic time series, which can be biased by changes in behaviour (health-care seeking behaviour, test-seeking behaviour, etc) or in severity of the virus (e.g, due to vaccination). However, there were concerns that such data would be ill-suited for estimating $R(t)$ due to the presence of long-term shedders – individuals who continue to test positive for a long duration of time, following infection. \\
\indent
In general, an epidemic time series is linked to the time series of infection incidence by a convolution; if the function for the convolution is known then it is possible to estimate the time series of incidence, and from that $R(t)$ \cite{Huisman2022-nv}. However, in many instances the form of this function is not known accurately (or at all) and so $R(t)$ cannot be straightforwardly calculated. In these instances, it is common practice for the epidemic time series to be treated as a reasonable proxy for the infection incidence and estimates of $R(t)$ are obtained by assuming the same renewal process between downstream points in the natural history of infection, for example from onset to onset or from death to death \cite{Nash2022-ob}. Often this assumption is still made even when an estimate of the convolution function could be made \cite{Dighe2020-kq, Ali2013-yj, Shim2021-oa}; despite the existence of computer packages that allow the inclusion of a convolution function when estimating $R(t)$ \cite{Abbott2020_abc, Lytras_undated-jd}.\\
\indent
There are other potential biases in epidemic time series which may influence estimates of $R(t)$. However, biases introduced by assuming the epidemic time series to be a perfect proxy for infection incidence have often been overlooked. Here we investigate the form that these biases take and how they depend on the relationship between an epidemic time series and the infection incidence. We further investigate how estimates of $R(t)$ made during the SARS-CoV-2 pandemic may have been biased by the epidemic time series that were available (onset of symptoms, infection prevalence, hospitalisations and deaths considered), and how data could be better collected and used to minimise these biases.

\section{Background}
\subsection{The relationship between incidence and $R(t)$}
The reproduction number can be written as

\begin{equation}
    R = \int \eta(\tau) \text{d}\tau,
\end{equation}
where $\eta(\tau)$ describes the rate of secondary infections as a function of the time since a primary infection, $\tau$. Normalising $\eta(\tau)$ gives the generation time,

\begin{equation}
    g(\tau) = \frac{\eta(\tau)}{\int \eta(\tau) \text{d}\tau},
\end{equation}
the probability distribution describing the time between primary and secondary infections. The incidence at time $t$, $I(t)$ can then be written in terms of the incidence at times less than $t$:

\begin{align}\label{eqn_I}
    I(t) &= \int_{-\infty}^{0}I(t-\tau)\eta(\tau)\text{d}\tau     \\
    &= \int_{-\infty}^{0}I(t-\tau)g(\tau)R(t)\text{d}\tau.\nonumber
\end{align}
Where $R(t)$ is the reproduction number at time $t$. Expressed in this way the reproduction number at time t can be calculated directly from the time-series of incidence (if the generation time function is known --- which we will assume throughout) through the equation:

\begin{equation}
       R(t) =\frac{I(t)}{\int_{-\infty}^{0}I(t-\tau)g(\tau)\text{d}\tau}.
\end{equation}

\subsection{The relationship between epidemic time series and $R(t)$}
A general epidemic time-series, $J(t)$, can be written in terms of the incidence at times less than $t$ and a convolution function, $f(\tau)$, that describes the relationship between the two time series:

\begin{equation}\label{eqn_J}
    J(t) = \int_{-\infty}^{0}I(t-\tau)f(\tau)\text{d}\tau.
\end{equation}
The convolution function will take different forms depending on the epidemic time series. For example, for the time series of deaths due to an infectious disease, $f(\tau)$ would describe the average rate of deaths as a function of the time since infection. Substituting the relationship in equation \ref{eqn_I} into equation \ref{eqn_J} we obtain:

\begin{equation}\label{eqn_J2}
    J(t) = \int_{-\infty}^{0}\int_{-\infty}^{0}I(t-\tau-\zeta)g(\zeta)R(t-\tau)f(\tau)\text{d}\tau \text{d}\zeta .
\end{equation}
In general there is not a simple relationship between $J(t)$ and $R(t)$. To calculate $R(t)$ from $J(t)$ the convolution function, $f(\tau)$, would need to be known (and of course $g(\tau)$). However, it is often assumed that the same relationship exists between $J(t)$ and $R(t)$, as the one between $I(t)$ and $R(t)$:

\begin{equation}
     J(t) = \int_{-\infty}^{0}J(t-\zeta)g(\zeta)R_A(t)\text{d}\zeta.
\end{equation}
Here $R_A(t)$ is the apparent reproduction number inferred from a general epidemic time series under this incorrect assumption, and is given by:

\begin{equation}\label{eqnR_A}
    R_A(t) = \frac{J(t)}{ \int_{-\infty}^{0}J(t-\zeta)g(\zeta)\text{d}\zeta}.
\end{equation}
In general $R_A(t) \neq R(t)$, but under certain conditions relationships between the two will hold.

\subsubsection*{$R(t) = $ constant}
If $R(t)$, has been constant for the period of time over which $f(\tau)$ goes to $0$ then equation \ref{eqn_J2} can be written as:

\begin{equation}
        J(t) = \int_{-\infty}^{0}J(t-\zeta)g(\zeta)\overline{R(t)}\text{d}\zeta.
\end{equation}
Where we have written $R(t)$ as $\overline{R(t)}$ to make it clear this equation is only valid when $R(t)$ is constant. In this situation the relationship between $J(t)$ and $R(t)$ is the same as the relationship between $I(t)$ and $R(t)$ and we have

\begin{equation}
    R_A(t) = \overline{R(t)}.
\end{equation}
\subsubsection*{$f(\tau) = \delta(\mu-\tau)$}
If the convolution function, $f(\tau)$, is the Dirac-delta function with mean $\mu$, $\delta(\mu-\tau)$, then $J(t)$ can be written as:

\begin{equation}
    J(t) = I(t-\mu).
\end{equation}
$R_A(t)$ can then be written as:

\begin{align} \label{eqn_lag}
    R_A(t) &= \frac{J(t)}{ \int_{-\infty}^{0}J(t-\zeta)g(\zeta)\text{d}\zeta}\\
    &=  \frac{I(t-\mu)}{ \int_{-\infty}^{0}I(t-\mu-\zeta)g(\zeta)\text{d}\zeta}\nonumber \\
    &= R(t-\mu)\nonumber.
\end{align}

\section{Results}
\subsection{The effect of convolution on estimates of $R(t)$}
\begin{figure}
\centering
\includegraphics[height= 0.7 \textheight]{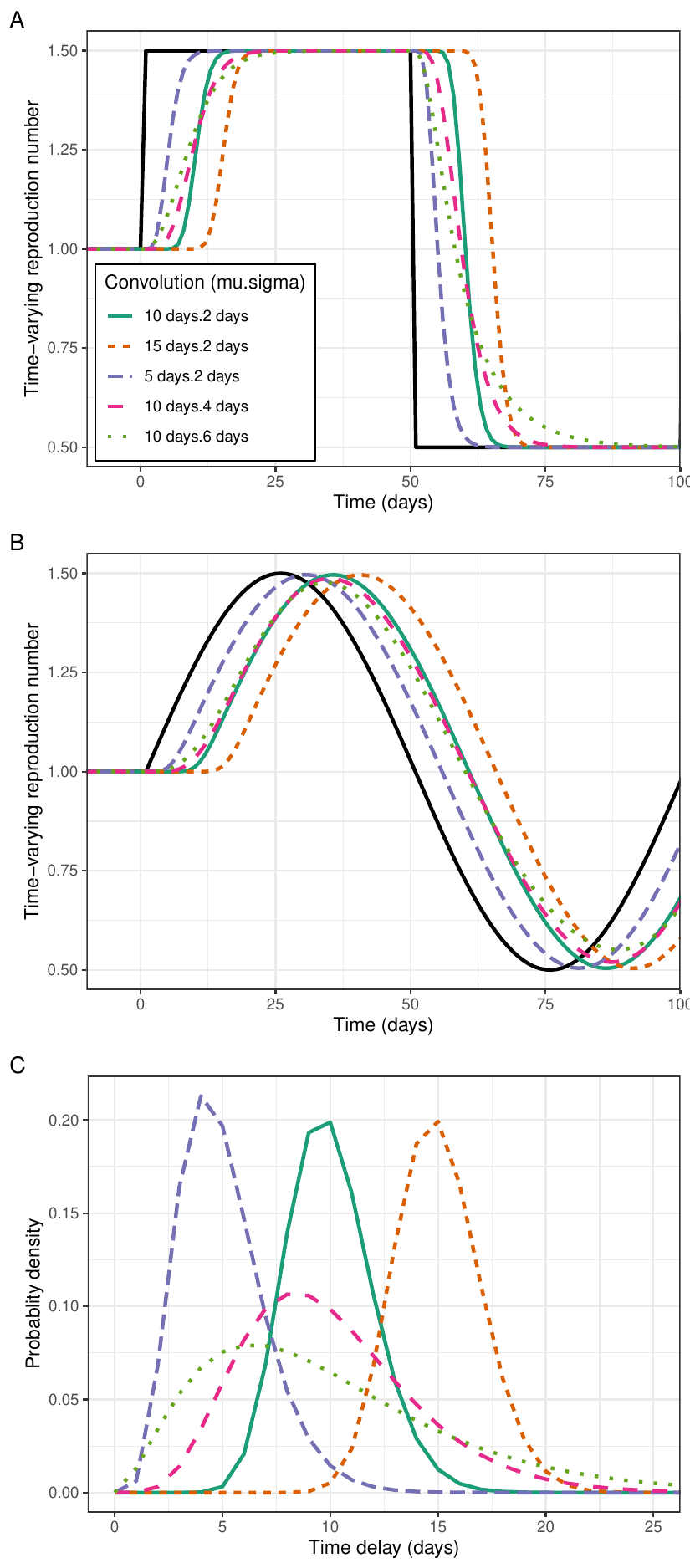}
\caption{{\bf Estimates of the time-varying reproduction number from epidemic time series with different gamma distributed convolution functions.}
(A, B) The actual reproduction number, $R(t)$ (black), and the apparent reproduction number, $R_A(t)$ (coloured), for epidemic time series with different gamma distributed convolution functions. (C) The gamma distributed convolution functions used in (A) and (B).
}
\label{fig1}
\end{figure}

\begin{figure}
\includegraphics[width=\textwidth]{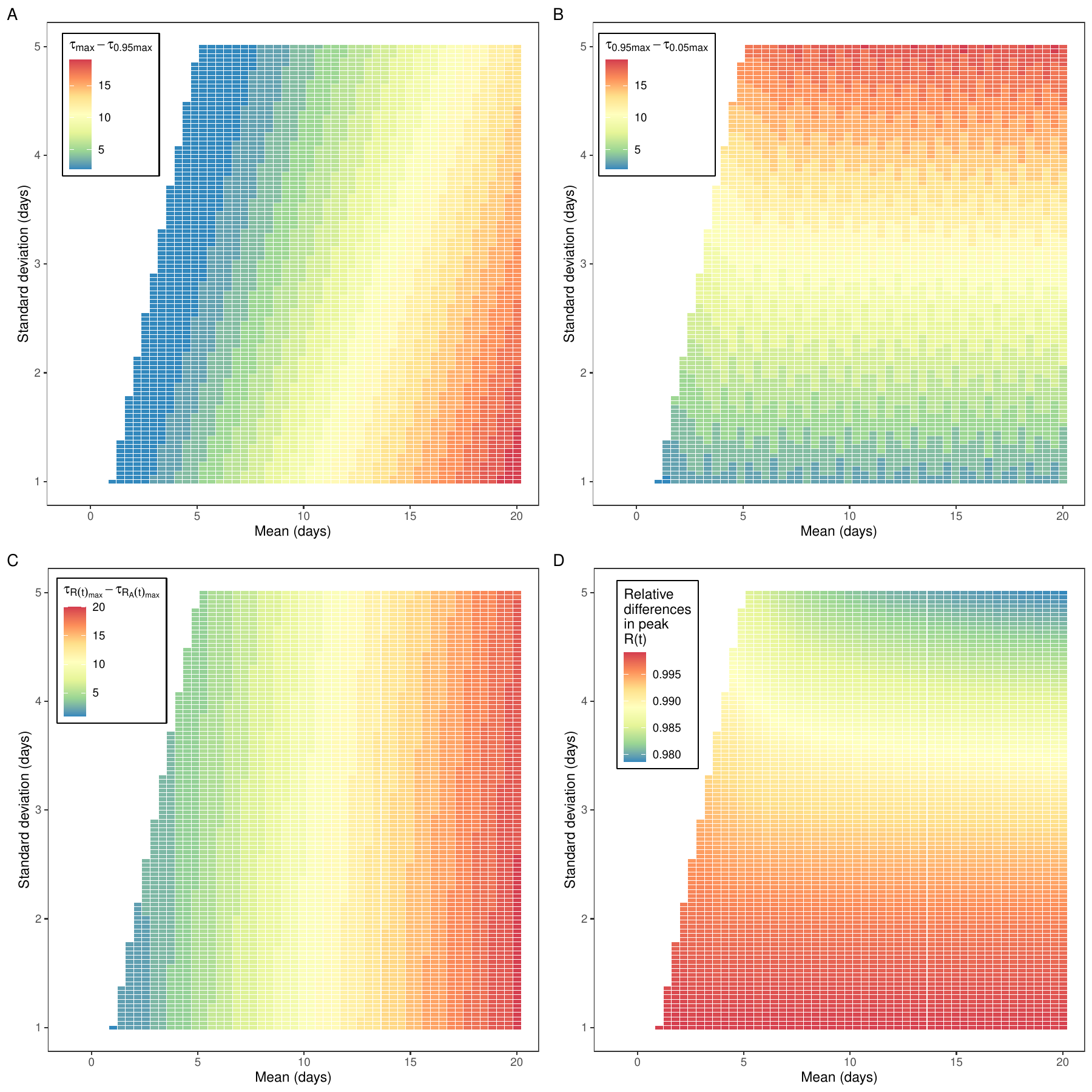}
\caption{{\bf Diagnostic differences between $R(t)$ and $R_A(t)$.}
Diagnostic differences between $R(t)$ and $R_A(t)$, for different gamma-distributed convolution functions. Only convolution functions with a mean value greater than their standard deviation are included. (A,B) Differences between $R_A(t)$ and $R(t)$, when $R(t)$ describes a step decrease (see Figure \ref{fig1}a, day 50). (A) The number of days it takes for $R_A(t)$ to fall below 95\% of its maximum value following the step change in $R(t)$. (B) The number of days between $R_A(t)$ falling below 95\% of its maximum value and falling below 5\% of its maximum value. (C,D) Differences between $R_A(t)$ and $R(t)$, when $R(t)$ describes a sine wave (see Figure \ref{fig1}b). (C) The number of days between $R(t)$ reaching its maximum value and $R_A(t)$ reaching its maximum value. (D) A measure of the difference between the peak values of $R(t)$ and $R_A(t)$, defined as $1-\frac{max(R(t)) - max(R_A(t))}{max(R(t)) - min(R(t))}$ it reflects the proportion of the maximum value that $R_A(t)$ reaches.
}
\label{fig2}
\end{figure}

Trends in the apparent reproduction number, $R_A(t)$, lagged trends in the actual reproduction number, $R(t)$, in ways that were dependent on the shape of the underlying infection time series. Infection incidence was simulated for two time series of $R(t)$: sinusoidal and square waves. A sinusoidal wave describes gradual changes in $R(t)$; gradual changes in $R(t)$ would be expected over the course of an epidemic (due to the depletion of susceptibles). A square wave describes rapid changes in $R(t)$, which would be expected when some public health interventions (e.g., lockdowns, school closures) are implemented. Epidemic time series and their associated $R_A(t)$ were then calculated for different gamma distributed convolution functions (Figure \ref{fig1}). As we would expect, the greater the mean of the convolution function, the greater the lag between $R_A(t)$ and $R(t)$. When $R(t)$ changed gradually (simulated as a sine function in time) (Figure \ref{fig1}b) there was a delay between $R(t)$ and $R_A(t)$ reaching their respective maximums (Figure \ref{fig2}c). This time delay was approximately equal to the mean of the convolution function. When $R(t)$ underwent a step decrease (figure \ref{fig1}a) there was a delay between $R(t)$ decreasing and $R_A(t)$ decreasing (with similar results for a step increase). Measuring the time delay between $R(t)$ decreasing (a step change) and $R_A(t)$ falling to under 95\% of its maximum value (Figure \ref{fig2}a) we observed a greater delay for convolution functions with greater means and lower standard deviations.\\
\indent
In addition to $R_A(t)$ lagging $R(t)$, trends in $R_A(t)$ (the shape of the curve over time) were distorted relative to $R(t)$, with increases in the standard deviation of the convolution function resulting in greater levels of distortion. When $R(t)$ changed gradually (simulated as a sine function in time) the maximum value of $R_A(t)$ was lower than the maximum value of $R(t)$. The greater the standard deviation of the convolution function the greater the relative reduction in the maximum value of $R_A(t)$ (Figure \ref{fig2}d). When $R(t)$ underwent a sharp decrease (a step change over a single day) $R_A(t)$ decreased at a slower rate (Figure \ref{fig1}a). Measuring the time delay between $R_A(t)$ falling from 95\% of its maximum value to 5\% of its maximum value we observed a greater delay when the standard deviation of the convolution function was greater (Figure \ref{fig2}b).

\subsection{Case study: SARS-CoV-2 epidemic time series}
\begin{figure}
\includegraphics[width=\textwidth]{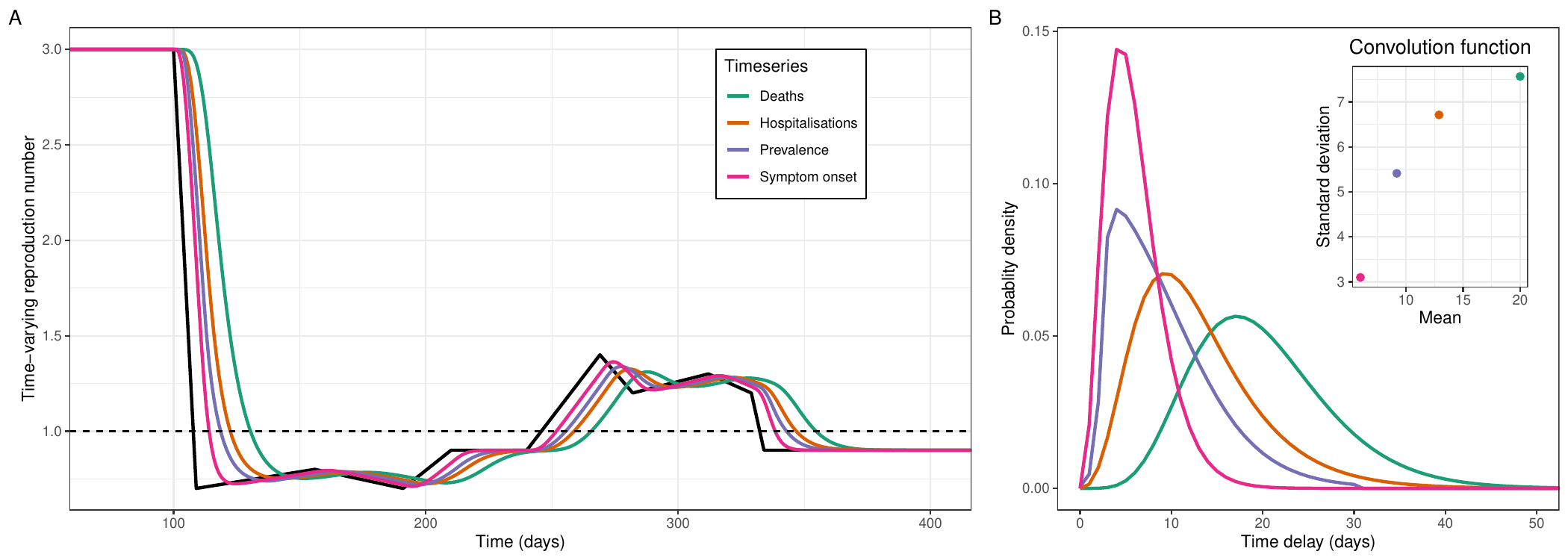}
\caption{{\bf Trends in $R(t)$ and $R_A(t)$ for representative epidemic time series during the SARS-CoV-2 pandemic.}
(A) The actual reproduction number, $R(t)$ (black), and the apparent reproduction number, $R_A(t)$ (coloured), for epidemic time series with different convolution functions. $R(t)$ was chosen to resemble a possible trajectory of $R(t)$ in England from May 2020 to December 2020 (B) The probability distribution of the convolution functions for deaths (Green), hospitalisations (Orange), prevalence (Purple), symptom onset (Pink). The mean and standard deviation of all convolution functions are presented in the inset graph.
}
\label{fig3}
\end{figure}

For all epidemic time series considered (that would have been available during the SARS-CoV-2 pandemic) there were clear differences between $R_A(t)$ and $R(t)$, during a representative scenario (chosen to resemble trends in $R(t)$ for SARS-CoV-2 in England from March to December 2020) (Figure \ref{fig3}). During periods when $R(t)$ was constant, the value of $R_A(t)$ was the same (as expected). When $R(t)$ changed, $R_A(t)$ lagged $R(t)$ and had its shape distorted in a way that was dependent on the shape of the convolution function.\\ 
\indent
The values of $R_A(t)$ calculated from the time series of symptom onset most closely resembled the underlying $R(t)$. The mean and standard deviation of the convolution function for symptom onset was lower than all other time series considered (Figure \ref{fig3}b). This was expected for deaths and hospitalisations, for which the convolution functions were composed of the distributions of time from infection to symptom onset (incubation period) and time from symptom onset to outcome (death/hospitalisation) (see Methods). Interestingly, the peak probability density of the convolution function for infection prevalence was the same as for symptom onset, but due to a long tail in the distribution the mean and standard deviation were greater.

\subsection{Improving inference using frequent testing}
\begin{figure}
\includegraphics[width=\textwidth]{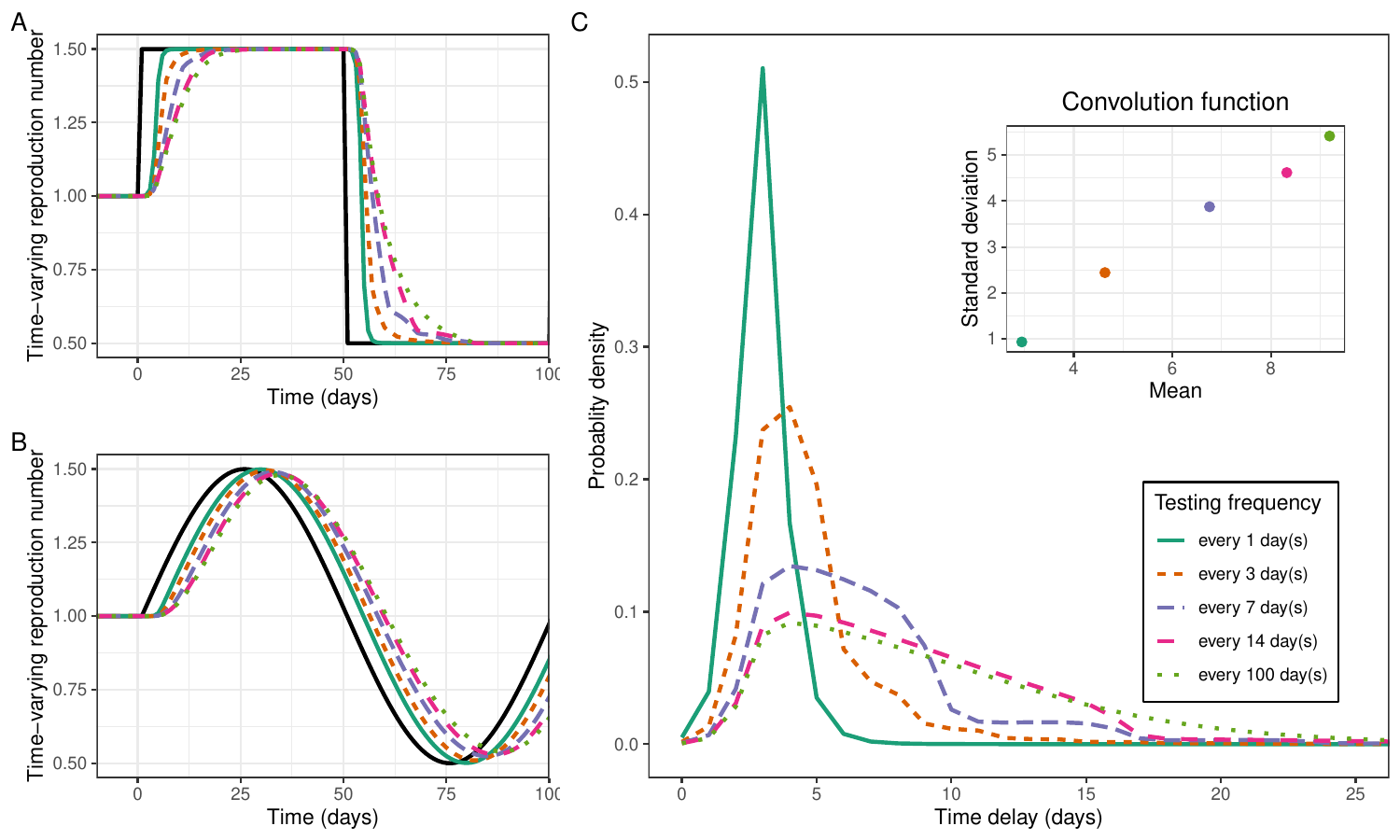}
\caption{{\bf The effect of increasing testing frequency on estimates of $R_A(t)$.}
(A,B) The actual reproduction number, $R(t)$ (black), and the apparent reproduction number, $R_A(t)$ (coloured), for epidemic time series with different convolution functions. (C) The probability distribution of convolution functions for different frequencies of rt-PCR testing. Testing frequency of 'every 100 days' reflects no repeat testing and has the same convolution function as for prevalence in figure \ref{fig3}. The mean and standard deviation of all convolution functions are presented in the inset graph.
}
\label{fig4}
\end{figure}

Increasing the frequency of random/asymptomatic testing reduced the mean and standard deviation of the convolution function (Figure \ref{fig4}c). Random testing or asymptomatic testing for SARS-CoV-2 should, in the absence of repeat testing, have a convolution function describing the probability of testing positive following infection. In our simulations, increasing the frequency of testing decreased the mean convolution function, as infections were more likely to be identified earlier. Testing once every three days gave a convolution function with a lower mean than that of symptom onset. Testing every day resulted in a convolution function with mean and standard deviation of approximately 3 days and 1 day respectively. The lower mean and standard deviation of the convolution function resulted in $R_A(t)$ estimates more closely resembling $R(t)$ as expected (Figures \ref{fig4}a and \ref{fig4}b). 

\subsection{Improving inference from infection prevalence studies}
\begin{figure}
\includegraphics[width=\textwidth]{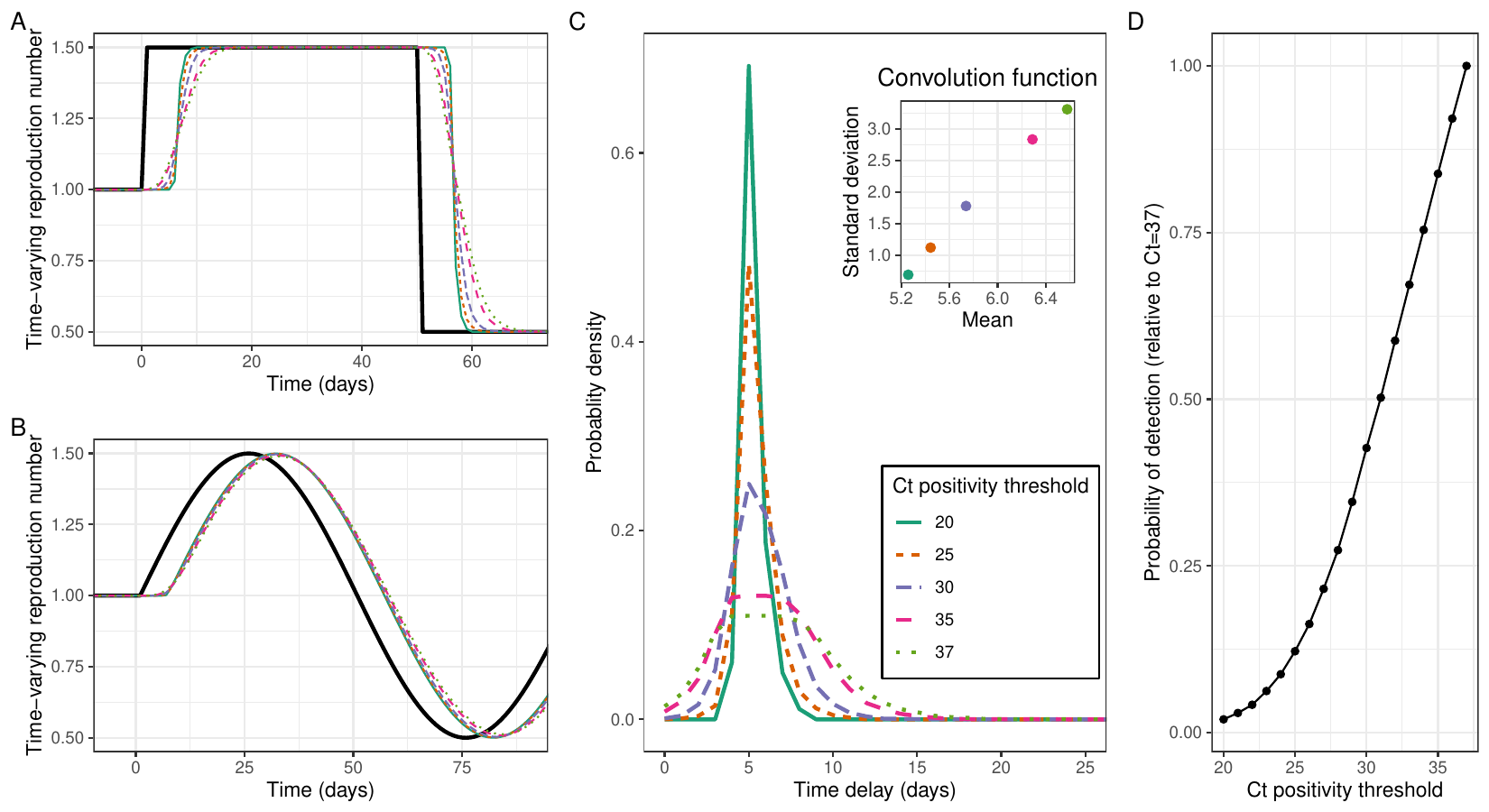}
\caption{{\bf The effect of imposing stricter Ct thresholds for positivity on estimates of $R_A(t)$.}
(A,B) The actual reproduction number, $R(t)$ (black), and the apparent reproduction number, $R_A(t)$ (coloured), for epidemic time series with different convolution functions. (C) The probability distribution of convolution functions for different Ct thresholds of rt-PCR positivity. Note that the convolution function was defined from a model of population Ct values rather that rt-PCR positivity, and so will not match the convolution function for prevalence in figure \ref{fig3}. The mean and standard deviation of all convolution functions are presented in the inset graph. (D) The proportion of individual tests classified as positive relative to the number defined as positive with a Ct threshold of 37. 
}
\label{fig5}
\end{figure}

Imposing a stricter viral threshold for positivity reduced the standard deviation of the convolution function (Figure \ref{fig5}c). rt-PCR tests, which are often used to test for the presence of SARS-CoV-2, can also provide information on the Ct value (proxy for viral load) of a positive test. By only including individual tests with a Ct value lower (higher viral load) than a certain threshold, it becomes less likely that early-infections or long-term shedders are included within the epidemic time series. This led to substantial decreases in the standard deviation of the convolution function, and smaller decreases in the mean of the convolution function. Accordingly, this led to the shape of $R_A(t)$ more closely resembling that of $R(t)$, while the lag between the two signals stayed approximately constant (Figure \ref{fig5}a and \ref{fig5}b). Imposing a stricter threshold reduced the number of individuals classified as positive and therefore the power of any analysis (Figure \ref{fig5}d).

\section{Discussion}
During the SARS-CoV-2 pandemic, estimates of the time-varying reproduction number, $R(t)$, were crucial for informing public health interventions. In most counties there were numerous epidemic time series available for estimating $R(t)$. We have described the biases that will be present in estimates of ‘$R(t)$’ when it is assumed that these time series are reliable proxies for infection incidence (and not the convolution of infection incidence with some other function). This assumption is commonly made \cite{Nash2022-ob}, despite methods existing to incorporate estimates of the convolution function into estimates of $R(t)$ \cite{Abbott2020_abc}. We presented the convolution functions representing four different epidemic time series for SARS-CoV-2: date of symptom onset (symptom reporting studies), prevalence (rt-PCR testing of random samples or asymptomatic testing), hospitalisations and deaths. Additionally, the convolution function for case-numbers, obtained from mass testing (available in many countries worldwide), would likely be a linear combination of the convolution functions for prevalence (asymptomatic testing) and symptom onset (symptomatic testing) with an additional delay to test distribution (delay between symptom onset and positive test).\\
\indent
If the exact form of these convolution function was known, then estimates of $R(t)$ could easily be made from the corresponding epidemic time series \cite{Huisman2022-nv}, but this is not always the case. During the early stages of a pandemic there is little epidemic information, and so convolution functions cannot be inferred. During later periods of a pandemic there can still be great uncertainty in estimates of the convolution function. There is also no guarantee that the convolution functions remain constant over time. For example, rates of symptomatic infection can change (effecting mass testing) \cite{Eales2022-vf, Eales2022-if}, severity can change (effecting hospitalisation and death time series) \cite{Eales2023-wm}, and even the duration an individual remains test positive can change due to new variants \cite{Eales2022-aq}, or due to vaccination \cite{Kissler2021-ng}. Thus, there are times when only the apparent reproduction number, $R_A(t)$, can be estimated, and so understanding the inherent biases in these estimates is crucial.\\
\indent
Changes in $R_A(t)$ lagged changes in $R(t)$, with the duration of the lag controlled by the mean of the convolution function. If timeliness in $R(t)$ estimates are required (e.g., for pandemic surveillance), then this time lag is a major limitation (though estimates could still be useful for retrospective analysis). The temporal trends in $R_A(t)$ were distorted relative to $R(t)$, with a greater degree of distortion when variance in the convolution function was greater; If there was no variance in the convolution function (standard deviation=0), $R_A(t)$ would exhibit the exact same temporal trends as $R(t)$ (lagged by the mean of the convolution function). This distortion was most pronounced when $R(t)$ changed rapidly, and less pronounced when $R(t)$ changed more gradually. During many epidemics only gradual changes in $R(t)$ are observed (e.g., due to the changing proportion of susceptibility as individuals are infected and develop immunity) and such distortions would be minimal. However, in some instances $R(t)$ may change rapidly, such as when restrictions are implemented \cite{Knock2021-ug}  --- for example lock downs during the SARS-CoV-2 pandemic --- or when schools are closed/opened \cite{Li2021-uz}. In these instances, there will be a high degree of distortion between $R(t)$ and $R_A(t)$.\\
\indent
The convolution function for SARS-CoV-2 symptom onset had the smallest mean and variance, of the four convolution functions considered, which would result in estimates of $R_A(t)$ that were the most similar to $R(t)$. However, there are often multiple circulating pathogens which may present similar symptom profiles. To estimate $R_A(t)$, it must be determined which individuals reporting symptoms are infected with the relevant pathogen. For this symptomatic testing the exact convolution function would rely on a distribution describing the time between symptom onset and testing (testing delay); this would likely increase the convolution function's mean and variance. To improve estimates of $R_A(t)$ the testing delay distribution should be reduced as much as possible. Additionally, the date of symptom onset should be routinely collected; this would reduce the mean and variance of the convolution function (no testing delay component required) and would also allow $R(t)$ to be estimated (once the incubation period has been estimated) as the convolution function would be less likely to vary over time (testing delay can change over time).\\
\indent
Random sampling has been proposed as a method for tracking COVID-19 infections \cite{Dean2022-ub}, without the biases present in standard pandemic surveillance methods such as mass testing \cite{Ricoca_Peixoto2020-wj}. There has been some concern that such studies would be less well suited for estimating $R(t)$, due to the presence of long-term shedders --- individuals who continue to test positive for a long duration of time, following infection. Random sampling studies measure the infection prevalence, the proportion of a population that test positive for the virus. We have demonstrated that the convolution function for prevalence of SARS-CoV-2 has a mean and variance only slightly greater than that of symptom onset. In fact, compared to mass-testing, which relies heavily on symptomatic testing, it is likely that the mean and variance of the convolution functions would be comparable, depending on the testing delay distribution.\\
\indent
The convolution function for symptomatic testing can at best match the incubation period of the pathogen. In contrast we have shown that there exist methods for improving the inference of $R_A(t)$ based on prevalence studies. Imposing a stricter threshold for positivity can reduce variance of the convolution function (though larger samples may be required as more positive tests would be excluded). It is worth noting that the Ct value model we presented is not a perfect analogue to rt-PCR positivity and so results are informative, but not the exact outcome that would be obtained using a stricter threshold for positivity. Increasing the frequency of testing could also reduce both the mean and variation of the convolution function. We observed that for daily testing the convolution function’s mean and variance was smaller than that of the incubation period of SARS-CoV-2. It is highly unlikely that a random sampling study, which is very expensive, could be designed to undertake such high rates of testing. However, the convolution function presented would also be valid for asymptomatic testing. Regular asymptomatic testing was performed by many individuals and businesses throughout the pandemic. Studies or surveillance systems set up around frequent asymptomatic testing could result in estimates of $R_A(t)$ that more closely resemble $R(t)$. In the latter stages of the pandemic Lateral Flow Viral Antigen detection devices (LFDs) were regularly used by some individuals. LFDs are less sensitive in general than rt-PCR testing, but they are still highly sensitive at detecting infections with high viral loads \cite{Joint_PHE_Porton_Down_University_of_Oxford_SARS-CoV-2_test_development_and_validation_cell2020-bg}. This is at times a limitation as less infected individuals will be identified, but the resulting convolution function would likely have even less variation.\\
\indent
We have only considered the biases in $R(t)$ due to the convolution function. There are often many other sources of bias associated with epidemic time series data \cite{Ricoca_Peixoto2020-wj}. For example, changing testing rates can bias estimates of $R(t)$ obtained from cases identified through mass testing \cite{Eales2022-yr}. Additionally, estimates of $R(t)$ also rely on accurate estimates of the generation time distribution \cite{Wallinga2007-qv}. The generation time is not always well characterised (especially during the early period of a pandemic) and in some circumstances it can change over time \cite{Abbott2022-ku}. When the generation time is not known (or poorly estimated), the growth rate is regularly used to quantify epidemic growth and decline as an alternative to $R(t)$ \cite{Parag2022-jw}. However, though the growth rate is independent of generation time, estimating it from a general epidemic time series would introduce a similar set of biases as for $R(t)$ due to the convolution function.

\section{Conclusion}
There are clear biases when estimating $R(t)$ from an epidemic time series, under the assumption that the time series is a perfect proxy for infection incidence. By collecting data to inform on the convolution function linking the epidemic time series to incidence, estimates of $R(t)$ could be made without such biases. During periods in which the convolution function cannot be estimated, the apparent reproduction number can and should still be estimated. Despite its biases, it is highly useful for quantifying transmission dynamics and informing public health interventions. To minimise the inherent biases present, more careful consideration should be given into what epidemic data is used and how it can be better collected and manipulated. Additionally, studies in which it is the apparent reproduction number being estimated should state this explicitly, making it clear what possible biases may be present.

\section{Materials and Methods}

\subsection{Simulating epidemic time series}
The time series of $R(t)$ are pre-specified. $R(t)$ is assumed to be 1 for a period of 100 days (days -99 to 0) prior to the simulation for ease of later integral calculations (integration's are all performed over the previous 100 days). The time series of $R(t)$ can then be defined in any way. We define two main time series of $R(t)$ following a square wave and a sine wave, both with a period of 100 days, a maximum value of 1.5 and a minimum value of 0.5. We also define a time series of $R(t)$ that follows approximately the same trend as the estimated $R(t)$ in England from March 2020 to December 2020 \cite{Knock2021-ug}. With a time series of $R(t)$, the time series for infection incidence can be calculated through equation \ref{eqn_I}:

\begin{equation}
    I(t) = \int_{-\infty}^{0}I(t-\tau)g(\tau)R(t)\text{d}\tau.
\end{equation}
The generation time is assumed to be gamma distributed with a mean of 6.36 days and a standard deviation of 4.20 days \cite{Bi2020-ls}. This was chosen to match the generation time of SARS-CoV-2, but as we assume perfect knowledge of the generation time all results are independent of the choice of the generation time distribution used. When calculating the value of $I(t)$ we perform the integral over the previous 100 days. This ensures the integral has fallen to approximately 0. Days -99 to 0 are assumed to have $I(t)=1$. \\
\indent
General epidemic time series, $J(t)$ are calculated directly from $I(t)$ through equation \ref{eqn_J}:

\begin{equation}
    J(t) = \int_{-\infty}^{0}I(t-\tau)f(\tau)\text{d}\tau.
\end{equation}
The integration is again performed over the previous 100 days, ensuring the integral has fallen to approximately 0. $f(\tau)$ is a convolution function which we define over a period of only 100 days.

\subsection{Calculating the reproduction number}
We calculate the apparent reproduction number, $R_A(t)$, directly from the general epidemic time series $J(t)$ through equation \ref{eqnR_A}:

\begin{equation}
    R_A(t) = \frac{J(t)}{ \int_{-\infty}^{0}J(t-\zeta)g(\zeta)\text{d}\zeta}.
\end{equation}
The integration is again performed over the previous 100 days, ensuring the integral has fallen to approximately 0. As before $g(\zeta)$ is the generation time distribution which is known exactly when performing the calculation.

\subsection{Convolution functions}
We define many convolution functions, $f(\tau)$, for converting $I(t)$ into a general epidemic time-series. All convolution functions are defined only for values of $\tau$ between 0 and 100.

\subsubsection*{Gamma-distributed convolution functions}
The gamma-distributed convolution functions are defined using the gamma distribution. A gamma distribution defined with scale parameter, $k$, and shape parameter, $\theta$ has mean, $k\theta$ and standard deviation, $k^{1/2}\theta$. We then define our gamma-distributed convolution functions with mean, $\mu$ and standard deviation, $\sigma$ using the inverse relationships: $k=\mu^2/\sigma^2$ and $\theta=\sigma^2/\mu$.

\subsubsection*{Convolution functions for SARS-CoV-2 epidemic time series}
The convolution function for symptom onset is defined by the incubation period of SARS-CoV-2, which we assume is gamma distributed with mean 6.0 days and standard deviation 3.1 days \cite{Linton2020-oi}.\\
\indent
The convolution functions for deaths and hospitalisations are calculated using the equation:

\begin{equation}
    f(\tau) = \int_{\tau}^{0} f_1(\tau-\zeta) f_2(\zeta) \text{d}\zeta.
\end{equation}
In this equation $f_1(\tau)$ describes the incubation period of SARS-CoV-2 (same gamma distribution as above) and $f_2(\zeta)$ describes the distribution of the time between symptom onset and death/hospitalisation. For deaths we assume $f_2(\zeta)$ is gamma distributed with mean 15.0 days and standard deviation 6.9 days \cite{Linton2020-oi}. For hospitalisations we assume $f_2(\zeta)$ is gamma distributed with mean 7.8 days and standard deviation 6.0 days \cite{Hawryluk2020-qf}. The parameters used for all distributions are only meant to be informative, there are many other potential values that could have been selected from the literature.\\
\indent
The convolution function for prevalence is defined using the probability of testing positive as a function of time since infection. We use the median of the modelled probability of testing positive (as a function of time since infection) estimated in Hellewell et al 2021 \cite{Hellewell2021-fo}.

\subsubsection*{Convolution function for frequent testing}
The convolution function for frequent testing is calculated using the distribution describing the probability of testing positive as a function of time since infection (the convolution function for prevalence). When there is frequent testing the convolution function describes the day in which an individual first tests positive only. Writing the probability of testing positive $\tau$ days after infection as $P(\tau)$, we can calculate the probability of first testing positive $\tau$ days after infection as:

\begin{equation}
    P(\text{first positive}, \tau) = P(\tau) \prod_{n=1}^{N-1} 1-P(\tau-n\times \gamma),
\end{equation}
where $N$ is the total number of tests performed since infection and $\gamma$ is the number of days between tests. The product in the above equation is calculating the probability of testing negative in all previous tests.

\subsubsection*{Convolution function for different Ct value thresholds}
The convolution function for different Ct value thresholds of positivity was calculated by simulating the trajectory of Ct values over time since infection for 1000 individuals. The proportion of individuals with a Ct value less than a threshold Ct value was then calculated as a function of time since infection. Ct values were simulated using the results and data of Hay et al 2022 \cite{Hay2022-dn} (Ct values represent ORF1ab). In the paper an individual's Ct value trajectories can be quantified using three parameters: the minimum Ct value reached, the time taken for Ct value to decrease from the limit of detection (Ct value=40) to its minimum value (linear decrease assumed), and the time taken for Ct value to increase from its minimum value to the limit of detection (linear increase assumed). The parameters for an individual are assumed to be drawn from parameter distributions describing the population as a whole. To simulate 1000 individuals we extracted 1000 individual level parameters from the population parameter distributions estimated for the model fit to the data for all Omicron infections. We had to assume the time of infection for all individuals. We assumed that all individuals reached peak viral load (minimum Ct value) exactly 5 days after infection.

\backmatter

\section*{Declarations}

\subsection*{Conflicts of interest}
The authors declare no competing interests.

\bibliography{sn-bibliography}



\end{document}